\documentclass[11pt,english,amsmath]{iopart}
\usepackage{iopams}
\usepackage{color}
\usepackage{graphicx}
\usepackage[official]{eurosym}
\usepackage{setstack}
\usepackage{cite}

\newcounter{XXXcounter}

\newcommand{\be}{\begin{equation}}
\newcommand{\ee}{\end{equation}}
\newcommand{\bd}{\begin{displaymath}}
\newcommand{\ed}{\end{displaymath}}
\newcommand{\BE}{\begin{eqnarray}}
\newcommand{\EE}{\end{eqnarray}}

\newcommand{\avg}[1]{\left\langle{#1}\right\rangle}

\newcommand{\boldeta}{\mbox{\boldmath $\eta$}}

\newcommand{\Q}{\mathcal{Q}}

\newcommand{\vr}{\varrho}
\newcommand{\eins}{1\!\!1}
\newcommand{\mean}[1]{\ensuremath{\langle{#1}\rangle}}

\newcommand{\QQ}{\mathcal{Q}}
\newcommand{\MM}{\mathcal{M}}
\newcommand{\HH}{\mathcal{H}}

\newcommand{\bra}[1]{\ensuremath{\langle#1|}}

\newcommand{\ket}[1]{\ensuremath{|#1\rangle}}

\newcommand{\ketbra}[1]{\ensuremath{| #1 \rangle \langle #1 |}}
\newcommand{\kommentar}[1]{}

\def\argmin{\mathop{\rm argmin}}
\def\argmax{\mathop{\rm argmax}}

\begin{document}

\title{Computing complexity measures for quantum
states based on exponential families}

\author{S\"onke Niekamp${}^{\mbox{\footnotesize\euro}}$, 
Tobias Galla${}^{\mbox{\footnotesize\pounds}}$, 
Matthias Kleinmann${}^{\mbox{\footnotesize\euro}}$, 
and
Otfried G\"uhne${}^{\mbox{\footnotesize\euro}}$}
\address{\euro~Naturwissenschaftlich-Technische Fakult\"at, 
Universit\"at Siegen, Walter-Flex-Stra{\ss}e~3, D-57068 Siegen, Germany}
\address{\pounds~Theoretical Physics, School of Physics and Astronomy, 
The University of Manchester, Manchester M13 9PL, United Kingdom} 
 
\begin{abstract}
Given a multiparticle quantum state, one may ask whether it can 
be represented as a thermal state of some Hamiltonian with 
$k$-particle interactions only. The distance from the exponential family defined by these thermal 
states can be considered as a measure of complexity of a given state. 
We investigate the resulting optimization problem and show how symmetries 
can be exploited to simplify the task of finding the nearest thermal state in a given exponential family. We also present 
an algorithm for the computation of the complexity measure and consider specific examples to demonstrate its applicability.
\end{abstract}

\ead{
\mailto{soenke.niekamp@uni-siegen.de}, 
\mailto{tobias.galla@manchester.ac.uk}, 
\mailto{matthias.kleinmann@uni-siegen.de},
\mailto{otfried.guehne@uni-siegen.de}}

\section{Introduction}
\label{sec:intro}
Understanding interacting multiparticle systems  is a central 
problem in many areas of physics, including condensed 
matter theory, quantum information processing, and complexity 
science. 
The difficulty of this problem arises from correlations between 
the particles: Typically, probability distributions of states of 
two or more particles do not factorize, hence the description of 
interacting systems requires significantly more 
parameters than the description of the same number of non-interacting 
particles. This makes the analysis of multiparticle systems challenging, 
but it also leads to novel and interesting phenomena such as quantum 
entanglement and classical complex behaviour. 

There are many approaches with which to characterize the complexity and 
correlations of multiparticle systems. In quantum information theory, 
much research has focused on entanglement measures \cite{plenio07, hororeview, 
tgreview}, but also on other forms of quantum correlations \cite{discordreview, cumulant}. 
Similarly, a number of different measures of complexity 
have been introduced for classical complex systems \cite{complexity1, complexity2, complexity3, complexity4, complexity5, complexity6}.

An approach which can be used to measure complexity in the classical 
as well as in the quantum domain makes use of interaction structures 
\cite{KOJA09, Zhou08, Zhou09, GaGu11}. In this approach one asks: Given 
a physical system, can its stationary state or density operator be viewed as a thermal state of an interacting 
particle system with $k$-body interactions only? If this is not the case, 
then how far is the state of the system from the space of all thermal states with  $k$-body 
interactions? This distance can then be used as a measure of complexity. 
To measure distances of this type it is useful to make use of the 
relative entropy, and 
of the underlying geometrical structure known in mathematics 
as {\it information geometry} \cite{amari01}. 
As already mentioned, an important feature of  this approach is the fact that 
it can be used for the quantum and the classical case: 
In classical multiparticle systems, each particle is assumed 
to be on one of a finite set of states at any one time, 
and the global physical 
system is described by a probability distribution over the products of 
the local states. In the spirit of Ref.~\cite{KOJA09} interactions 
are described by the Hamiltonian function 
on the state space, where a Hamiltonian is a $k$-particle Hamiltonian, if 
it is a sum of functions each of which acts on $k$-particles only. 
It is important to stress though that the Hamiltonian used 
in this approach is not necessarily a physical energy function, 
it is mostly a mathematical object with which to characterise 
the factorisation properties of the system's stationary state. 
Indeed, the approach is applicable to non-equilibrium systems 
as well, 
for which there may not be any energy function at all. In 
the quantum 
case, the physical system is described by a density matrix, 
and the Hamiltonian 
is an operator acting on the corresponding Hilbert space, but again it need not be that of an actual physical system.

A central problem for the practical application of this approach in the
quantum case is the calculation of the respective distances. 
For classical systems with binary states efficient 
algorithms are known \cite{CsSh04, cipi, GaGu11}. 
For the quantum case, some analytical results for 
states with a high degree of symmetry have been derived 
\cite{Zhou08}. Moreover, an algorithmic approach 
has been proposed in Ref.~\cite{Zhou09b}. Nevertheless 
the practical applicability of these ideas is not 
clear at present, and more general results and an 
overarching mathematical theory are as yet missing. 

In this paper, we present several results relating 
to the computation of the distance of a given
quantum state from the set of thermal states generated 
by Hamiltonians with $k$-particle interactions. 
First, we show how symmetries of the state can be 
used to simplify calculations of this type. Second, 
we present a new algorithm for the efficient 
computation of such distances. We also discuss how 
tools from convex 
optimization can be used in order to compute 
complexity measures for quantum
states.

Our paper is organized as follows: In Section 2, we introduce 
the relevant notation, in particular the formalism of quantum 
exponential families, and we explain the most relevant 
existing results. In Section 3 we show that for cases in which 
the quantum state being studied carries a certain symmetry 
the closest thermal state generated by $k$-particle Hamiltonians 
has the same symmetry. In Section 4
we present our algorithm and discuss its 
application to specific examples. In Section 5 
we discuss how results from optimization 
theory can be used to study this problem. 
Section 6 finally summarises our findings 
and we present an outlook on future lines of research.

\section{Exponential families of quantum states}
\label{sec:expfam}

In this section we introduce the theory of exponential families 
of interaction spaces for the special case of quantum states. 
We emphasize that most of the results presented here have 
been derived for the classical case by various authors  
(see, e.g., Refs.~\cite{amari01, KOJA09, AyKnauf07}) and for the quantum 
case mostly by D. L. Zhou \cite{Zhou08,Zhou09}. Nevertheless, since 
these results are rather scattered in the literature, we believe 
that a more comprehensive presentation can be useful.

\subsection{Exponential and Bloch representation}\label{sec:exp1}

We consider systems consisting of $n$ two-level systems 
(qubits) throughout; 
the generalization to higher-dimensional systems is 
straightforward. For later convenience, we first describe 
two different ways of representing quantum states of such systems.

The first possible representation is the \emph{exponential representation}.
It uses the fact that any quantum state can be considered as a thermal state
of some appropriately chosen Hamiltonian. More precisely, any $n$-qubit 
quantum state of full rank can be written as
\begin{equation}\label{eq:exprep}
\vr_{\rm exp}(\btheta)
=\exp\bigl(\sum_{\alpha_1,\ldots,\alpha_n}\theta_{\alpha_1,\ldots,\alpha_n}
\sigma_{\alpha_1}\otimes\cdots\otimes\sigma_{\alpha_n}\bigr)
\end{equation}
where the indices $\alpha_k$  run from $0$ to $3$ and the $\sigma_i$ are the Pauli 
matrices with the convention $\sigma_0 = \eins, \sigma_1 =\sigma_x$,
$\sigma_2=\sigma_y$ and $\sigma_3=\sigma_z$.
In the following, it will be convenient to use a multi-index notation,
\begin{equation}
  \vr_{\rm exp}(\btheta)
  =\exp\bigl(\sum_\alpha\theta_\alpha\sigma_\alpha\bigr), 
\end{equation}
where $\sigma_\alpha=\sigma_{\alpha_1}\otimes\cdots\otimes\sigma_{\alpha_n}$, 
and where $\alpha=(\alpha_1,\dots,\alpha_n)$.
The coefficient $\theta_{{0}}$ of the identity 
$\sigma_{{0}}=\eins^{\otimes n}$ in the above quantum states is not arbitrary, 
as it can be determined from the normalization 
condition $\tr\vr_{\rm exp}(\btheta)=1$. Explicitly one has
$\theta_{{0}}=-\psi(\btheta)$, where
\begin{equation}
  \label{eq:free}
  \psi(\btheta)=\ln\bigl\{\tr\bigl[\exp\bigl(\sum_{\alpha\neq 0}
  \theta_\alpha\sigma_\alpha\bigr)\bigr]\bigr\}.
\end{equation}
Any quantum state, $\vr$, of full rank can be written in the form of Eq. (\ref{eq:exprep}), and one can view the exponent in the exponential representation as 
a Hamiltonian of which $\vr$ is the thermal state. In this 
terminology, the function $\psi$ is up to a sign the free energy 
of statistical ensemble \cite{stingl, zia, balian}. It is important to note, however, 
that the Hamiltonian does not necessarily correspond to that of an actual 
physical system.

An alternative description of the quantum state is given by the {\it affine representation} 
or \emph{Bloch representation}. In this representation one writes the state as
\begin{equation}
\vr_{\rm aff}(\boldeta)
  =\frac{1}{2^n}\sum_\alpha\eta_\alpha\sigma_\alpha,
\end{equation}
where the coefficients are given by $\eta_\alpha= \tr(\vr_{\rm aff}\sigma_\alpha).$
Here, the normalization condition is simply given by $\eta_{{0}}=1$.
Note that in the affine representation the positivity 
of the density matrix results in additional restrictions 
on the coefficients $\boldeta$; these conditions, 
however, cannot normally be formulated straightforwardly \cite{kimura, byrd}.

We will now briefly discuss the connections between the 
two representations of quantum 
states. In order to do so consider two states 
$\vr$ and $\vr'$ of full rank in their
different representations, 
$\vr=\vr_{\rm exp}(\btheta)=\vr_{\rm aff}(\boldeta)$ 
and
$\vr'=\vr_{\rm exp}(\btheta')=\vr_{\rm aff}(\boldeta')$. 
Using the standard definition\footnote{Note that the binary 
logarithm is used in our definition of $D(\vr\Vert\chi)$.} 
of the relative entropy between two quantum states, $\vr$ and $\chi$,
\be
D(\vr\Vert\chi)=\tr[\vr\log_2(\vr)]-\tr[\vr\log_2(\chi)],
\ee
as well as the entropy $S(\vr)=-\tr(\vr\log_2\vr)$ of a single quantum state,
we then have
\begin{eqnarray}
\fl
\ln(2)D(\vr\Vert\vr')
&=-\ln(2)S[\vr_{\rm aff}(\boldeta)]
-\tr\Bigl\{\frac{1}{2^n}
\Big[\eins+\sum_{\alpha\neq 0}\eta_{\alpha}\sigma_\alpha\Big]
\Big[\sum_{\beta\neq 0}\theta'_\beta\sigma_\beta
    -\psi(\btheta')\eins\Big]\Big\}
    \nonumber
\\
\fl
&=\phi(\boldeta)+\psi(\btheta')
    -\sum_{\alpha\neq 0}\eta_\alpha\theta_\alpha',
\label{ddarstellung1}
\end{eqnarray}
where the function
$
\phi(\boldeta)=-\ln(2)S[\vr_{\rm aff}(\boldeta)]=-\ln(2)S(\vr)
$
is proportional to the entropy of $\vr$. With the scalar product
$
  \boldeta\cdot\btheta'
  =\sum_{\alpha\neq 0}\eta_\alpha\theta_\alpha'
$
this result takes the form
\begin{equation}
  \label{eq:quantcent}
  \ln(2)D(\vr\Vert\vr')
  =\phi(\boldeta)+\psi(\btheta')
  -\boldeta\cdot\btheta'.
\end{equation}
For the  special case in which $\vr=\vr'$ this reads
\begin{equation}
  \label{eq:quantlegendre}
  \phi(\boldeta)+\psi(\btheta)
  -\boldeta\cdot\btheta=0,
\end{equation}
a result which will become important below.
At this point it should be noted that the expression of 
Eq.~(\ref{eq:quantlegendre}) shows that $\psi(\btheta)$ 
and $\phi(\boldeta)$ are related by a Legendre transformation. 
More specifically, from Eq.~(\ref{eq:quantlegendre}) it follows
that $\eta_\alpha ={\partial\psi(\btheta)}/{\partial\theta_\alpha}$
and $\theta_\alpha={\partial\phi(\boldeta)}/{\partial\eta_\alpha}$
for all $\alpha\neq 0$. 

Similar structures are, of course, well known in statistical 
mechanics: a thermodynamic ensemble in statistical mechanics is defined by 
the requirement that some observables $A_i$ (e.g.~the Hamiltonian) have 
fixed expectation values (e.g.~the internal energy $U$). Maximizing the 
entropy of the statistical distribution of the ensemble under these 
constraints,  the thermal state of the ensemble comes out as 
$\vr \sim \exp\big(-\sum_i \lambda_i A_i\big)$, where the coefficients
$\lambda_i$ arise as Lagrange multipliers. It is then well-known that 
the Lagrange multipliers $\lambda_i$ are related to the expectation values 
$\mean{A_i}$ by a Legendre transformation (see page 40 in Ref.~\cite{stingl}, or \cite{zia, balian}).

Let us finally explain a useful theorem for the relative entropies between three
states. For the pairwise relative entropies of three full-rank states 
$\vr$, $\vr'$ and $\vr''$ one has
\BE 
\fl
D(\vr\Vert\vr'')  - D(\vr\Vert\vr')& - D(\vr'\Vert\vr'')
=D(\vr\Vert\vr'')-D(\vr\Vert\vr')-D(\vr'\Vert\vr'')
    +D(\vr'\Vert\vr')
    \nonumber 
    \\
    \fl
&={\frac{1}{\ln(2)}}
    \Bigl[\phi(\boldeta)+\psi(\btheta'')
    -\boldeta\cdot\btheta''
    -\phi(\boldeta)-\psi(\btheta')
    +\boldeta\cdot\btheta' 
    \nonumber \\
    &-\phi(\boldeta')-\psi(\btheta'')
    +\boldeta'\cdot\btheta''
    +\phi(\boldeta')+\psi(\btheta')
    -\boldeta'\cdot\btheta'\Bigr]
    \nonumber \\
    &={\frac{1}{\ln(2)}}
    (\boldeta-\boldeta')
    \cdot(\btheta'-\btheta'')
\EE
and thus
\begin{equation}
  \label{eq:quantpythagoras}
  D(\vr\Vert\vr'')=D(\vr\Vert\vr')+D(\vr'\Vert\vr'')
  +\frac{1}{\ln(2)}(\boldeta-\boldeta')
  \cdot(\btheta'-\btheta'').
\end{equation}
If the scalar product vanishes, this relation is also called 
the \emph{generalized Pythagoras theorem} \cite{amari01}.
  
\subsection{Exponential families and the information projection}\label{sec:exp2}

We next define the exponential families of states generated by 
$k$-particle Hamiltonians, objects of this type will be the 
focus of the work presented here.
For a given multi-index $\alpha=(\alpha_1,\dots,\alpha_n)$ 
we write $W(\alpha)$ for the weight of $\alpha$, i.e.,
the number of factors in the Pauli operator
$\sigma_\alpha=\sigma_{\alpha_1}\otimes\cdots\otimes\sigma_{\alpha_n}$ 
different from the identity. In other words the weight 
$W(\alpha)$ of a multi-index $\alpha$ is the number
of nonzero elements $\alpha_i$.

For any $1\le k\le n$ we then can define the so-called 
exponential family 
$\Q_k$ of thermal states of $k$-party Hamiltonians as
\be
  \Q_k=\bigl\{\vr |
  \vr=\exp\bigl(\sum_{\alpha:  W(\alpha)\le k}
  \theta_\alpha\sigma_\alpha\bigr)\bigr\},
\ee
so that $\QQ_k$ is the set of all quantum states, for which the 
exponential representation contains only $k$-body interactions in the 
Hamiltonian. Since the exponential representation is unique and the 
operators $\sigma_\alpha$ form a basis of the operator space, this definition
is unambiguous. It will also be useful to write $\HH_k$ for the 
space of all Hamiltonians containing only interaction terms up 
to weight $k$.

The set $\QQ_k$ represents a manifold in the space of all 
quantum states (see also Fig.~1), a direct characterization is 
not straightforward. Obviously, one has $\QQ_k \subset \QQ_{k+1}$ 
and consequently the exponential families define a hierarchy
\begin{equation}
  \Q_1 \subset\Q_2\subset\cdots\subset\Q_n,
\end{equation}
where $\Q_n$ is the set of all states with full rank and $\Q_1$ the set 
of all product states with full rank. For states $\vr$ with 
full rank we will then ask what the minimal order of 
interaction, $k$, is such that $\vr\in\Q_k$. For states which are 
not of full rank, the analogous question is whether the state 
is in the closure $\overline{\Q}_k$ of an exponential family. The introduction 
of the closure of exponential families makes the discussions and results 
that follow applicable to quantum states for which some of the eigenvalues 
vanish.

For a given state quantum state $\vr$ one can then construct the distance from the exponential family $\Q_k$, and in particular the state in $\Q_k$ which is the closest to $\vr$. This defines the so-called information
projection:

\noindent
{\bf Definition~1.}
The \emph{information projection} $\tilde{\vr}_k$ of a quantum 
state $\vr$ is the element of the exponential family $\Q_k$ which 
is the closest to $\vr$ with respect to the quantum relative entropy,
\begin{equation}
\tilde{\vr}_k=\mbox{argmin}_{\vr'\in\Q_k}D(\vr \Vert \vr'),
\end{equation}
where $D(\vr\Vert\chi)=\tr[\vr\log_2(\vr)]-\tr[\vr\log_2(\chi)].$ 
The distance to the information projection is then considered
as a complexity  or correlation measure and is given by
\be
\label{eq:qdist}
D_k(\vr)= \inf_{\vr' \in \Q_k} D(\vr\Vert\vr') =  D(\vr\Vert\tilde{\vr}_k)
\ee

\begin{figure}[t]
  \centering
  \def\svgwidth{0.4\textwidth}
  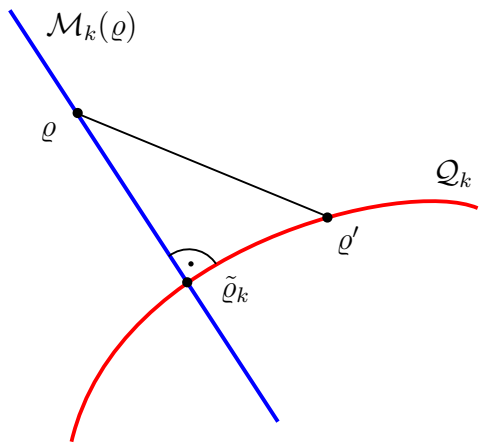
  \caption{\label{fig:q-projection}
  Illustration of the information projection onto a quantum exponential 
  family. Shown are the linear family $\MM_k(\vr)$ of distributions with 
  the same $k$-party reduced density matrices as $\vr$ (blue line), the 
  exponential family $\Q_k$ of thermal states of $k$-party Hamiltonians 
  (red curve) and the information projection $\tilde{\vr}_k$ of $\vr$ 
  onto $\Q_k$; and $\vr'$ represents an arbitrary state in $\Q_k$. See text for 
  further details.}
\end{figure}
 
In order to carry out the computation of this distance it is 
useful to have different characterizations of the information projection
$\tilde{\vr}_k.$ To this end, consider a given state $\vr$ and define
the set $\MM_k(\vr)$ of states with the same
$k$-party reduced density matrices as $\vr$,
\begin{equation}
  \MM_k(\vr)=\bigl\{\vr' \;\vert\; \vr_A'=\vr_A^{\phantom{\prime}}\
  \mbox{for all}\ A\subseteq\{1,\dots,n\}\ \mbox{with}\ |A|=k\big\},
\end{equation}
where $\vr_A=\tr_{\lbrace i_1,\ldots,i_n\rbrace\setminus A}(\vr)$
is the density matrix which is obtained from $\vr$ by tracing out all
qubits except those with indices in $A.$ We note that $\MM_k(\vr)$ is a linear subspace
of the space of all $n$-qubit density matrices, as opposed to the exponential
families $\QQ_k$.
Alternatively, one can also write
\begin{equation}
  \MM_k[\vr_{\rm aff}(\boldeta)]
  = \big\{\vr_{\rm aff}(\boldeta')
   \;\vert\;
  \eta_\alpha'=\eta_\alpha^{\phantom{\prime}}\
  \mbox{for all}\ \alpha\ \mbox{with}\
  W(\alpha)\le k\big\}.
\end{equation}
The following Lemma was first proven in Ref.~\cite{Zhou09} and
presents three equivalent constructions of the information 
projection. It also shows that the state $\tilde{\vr}_k$ in 
Definition 1 is unique.

\noindent
\\
{\bf Lemma 2.}
The following conditions on a quantum state $\tilde{\vr}_k$
are equivalent \cite{Zhou09}:
\\
(a) The state $\tilde{\vr}_k$ is the information projection in 
 the sense of Definition 1.
 \\
(b) The state $\tilde{\vr}_k$ is the maximizer of the von Neumann 
entropy in the set $\MM_k(\vr)$ of all states with the same 
$k$-party reduced density matrices as $\vr$,
\begin{equation}
\label{eq:quantmaxent}
\tilde{\vr}_k=\argmax_{\vr'\in \MM_k(\vr)}S(\vr').
\end{equation}
(c) The state $\tilde{\vr}_k$ is the unique element
of the intersection of the exponential family $\Q_k$ 
with the set $\MM_k(\vr)$ of all states sharing the 
same $k$-party reduced density matrices as $\vr$,
\begin{equation}
\label{eq:quantintersection}
\{\tilde{\vr}_k\}=\Q_k \cap \MM_k(\vr).
\end{equation}

The situation is illustrated in Fig.~\ref{fig:q-projection}. Instead of computing
$\tilde{\vr}_k$ by minimizing the distance from $\Q_k$, one can also look
at the intersection of $\Q_k$ with $\MM_k(\vr)$, or indeed maximize the
entropy among elements of the linear family $\MM_k(\vr)$.

{\it Proof.} Let us start with the case (b), and show that 
the characterisation (b) is equivalent to that of (a). The linear family 
$\MM_k(\vr)$ is defined as the set of all states $\vr'$ with the same 
$k$-particle reduced density matrices. This  condition is 
equivalent to the requirement that 
$\tr(\vr'\sigma_\alpha) = \lambda_\alpha = \tr(\vr\sigma_\alpha)$ 
for all $\alpha$ with $W(\alpha)\le k$. Now, maximizing the entropy
under the constraint of given expectation values is a well discussed
problem in statistical mechanics \cite{stingl}. The following results are 
known: The state maximizing the entropy is of the form 
$\tilde \vr_k  \sim \exp\{\sum_{\alpha: W(\alpha)\leq k} \theta_\alpha \sigma_\alpha\}$ 
and the maximum
is unique. So it is clear that the maximization in condition (b)
results in a unique state $\tilde \vr_k$ in $\QQ_k,$ we only have to show
that minimizes the relative entropy.

In order to show that, consider a third state $\vr'$ in $\QQ_k$ (see also 
Fig.~1). We apply Eq.~(\ref{eq:quantpythagoras}) to the state
$\vr=\vr_{{\rm aff}}(\boldeta)$, the state
$\tilde{\vr_k}=\vr_{\rm aff}(\tilde{\boldeta})
=\vr_{{\rm exp}}(\tilde{\btheta})$ [defined via Eq.~(\ref{eq:quantmaxent})]
and the states $\vr'=\vr_{\rm exp}(\btheta')$ in $\Q_k$, 
resulting in
\begin{equation}
D(\vr\Vert \vr')=D(\vr\Vert\tilde{\vr}_k)+D(\tilde{\vr}_k\Vert\vr')
    +\frac{1}{\ln(2)}
    (\boldeta-\tilde{\boldeta})
    \cdot(\tilde{\btheta}-\btheta').
\end{equation}
The terms in the scalar product with $W(\alpha)\le k$ vanish, as we have $\tilde \vr_k \in \MM_k(\vr)$ and thus 
$\eta_\alpha=\tilde{\eta}_\alpha$ for these $\alpha$. The terms with 
the terms with $W(\alpha)>k$ vanish because of 
$\tilde{\theta}_\alpha=\theta_\alpha'=0$. So one has 
$
D(\vr\Vert\vr')=D(\vr\Vert\tilde{\vr}_k)+D(\tilde{\vr}_k\Vert\vr'),
$
which implies that $\tilde{\vr}_k$, as defined in
Eq.~(\ref{eq:quantmaxent}), is also the unique state minimizing
the relative entropy $D(\vr\Vert\vr')$ among all $\vr'\in\Q_k$. 

Let us now turn to the characterisation (c). The state defined in (b) is obviously
in the intersection $\Q_k \cap \MM_k(\vr)$, so all we have to show
is that this intersection consists of only a single state. Let us assume
the contrary, so that there are two states $\vr_1$ and $\vr_2$
in $\Q_k \cap \MM_k(\vr).$ Then, applying 
Eq.~(\ref{eq:quantpythagoras}) and the same argument as above 
one finds that
$
0 = D(\vr_1 \Vert \vr_1)=D(\vr_1 \Vert \vr_2) + D(\vr_2 \Vert \vr_1).
$
Since the relative entropy is positive semidefinite, this implies that
$\vr_1=\vr_2.$
\hfill $\square$


\subsection{Complexity measures: Definitions and Properties}\label{sec:exp3}
As already mentioned, a central topic of this paper is the computation
of the distance $D_k(\vr)$ as defined in Eq.~(\ref{eq:qdist}). Before
presenting our results, it is useful to collect some of the properties 
of the distance measure $D_k(\cdot)$. First, note 
that $D_k$ can increase under local transformations, if $k \geq 2$ 
\cite{Zhou09, GaGu11}. This means that $D_k$ cannot in a naive way 
be viewed as a correlation measure, and so we prefer to call 
it a complexity measure.

Next one can define the \emph{degree of irreducible 
$k$-party interaction} as
\begin{equation}
\label{eq:qintmeas}
C_k(\vr)=D_{k-1}(\vr)-D_k(\vr),\qquad k=2,\ldots,n
\end{equation}
(where $D_n\equiv 0$). The quantity $C_k(\vr)$ describes the extent to which the 
approximation of a state, $\vr$, improves, if the allowed interactions in a Hamiltonian
increase from $(k-1)$-body interactions to $k$-body interactions.
By the generalized Pythagoras theorem, the last
definition is directly equivalent to
\begin{equation}
  \label{eq:qcrelent}
  C_k(\vr)=D(\tilde{\vr}_k\Vert\tilde{\vr}_{k-1}),\qquad k=2,\ldots,n-1.
\end{equation}
Furthermore, writing $\vr=\vr_{{\rm aff}}(\boldeta)$ and
$\tilde{\vr}_k=\vr_{{\rm aff}}(\tilde{\boldeta})
=\vr_{{\rm exp}}(\tilde{\btheta})$, we have, using
Eqs.~(\ref{eq:quantcent}) and (\ref{eq:quantlegendre}),
\BE
    \ln(2)D_k(\vr)
    &=\ln(2)D(\vr\Vert\tilde{\vr}_k)=
    -\ln(2)S(\vr)+\psi(\tilde{\btheta})
    -\boldeta\cdot\tilde{\btheta}
    \nonumber
    \\
    &=-\ln(2)S(\vr)+\psi(\tilde{\btheta})
    -\tilde{\boldeta}\cdot\tilde{\btheta}
    \nonumber
    \\
    &=-\ln(2)S(\vr)+\ln(2)S(\tilde{\vr}_k).
\EE
This shows that
\begin{equation}
  D_k(\vr)=S(\tilde{\vr}_k)-S(\vr),\qquad k=1,\ldots,n-1,
\end{equation}
and consequently
\begin{equation}
  \label{eq:qcdiffent}
  C_k(\vr)=S(\tilde{\vr}_{k-1})-S(\tilde{\vr}_k),\qquad k=2,\ldots,n-1.
\end{equation}
These different expressions for $D_k$ and $C_k$ can be useful 
for investigating the performance of numerical algorithms 
to compute the information projection: Having obtained the 
projections $\tilde{\varrho}_k$, one can compute the distances 
$D_k(\varrho)=D(\varrho\Vert\tilde{\varrho}_k)$ and the interaction 
measures $C_k(\varrho)$. The latter can be calculated in three 
different ways, namely via Eq.~(\ref{eq:qintmeas}), 
Eq.~(\ref{eq:qcrelent}) or Eq.~(\ref{eq:qcdiffent}).
If $\tilde{\varrho}_k$ is not the correct information projection, 
these three expressions will in general give different values.

Finally, let us discuss the case $k=1$ in some more detail. 
The quantity $D_1$ is also referred to as the 
{\em multi-information} \cite{AyKnauf07} or the
\emph{degree of total interaction}. It has an expansion 
into a telescopic sum of entropy differences
\begin{equation}
  C_{{\rm tot}}(\vr)=D_1(\vr)=\sum_{k=2}^nC_k(\vr).
\end{equation}
This is an orthogonal decomposition in the sense of the generalized Pythagoras
theorem.

The exponential family $\QQ_1$ consists of all product states 
(with full rank). The projection of a state $\vr$ onto this family 
is given by the tensor product of the one-party reduced density 
matrices,
\begin{equation}
  \tilde{\vr}_1=\vr_{\{1\}}\otimes\cdots\otimes\vr_{\{n\}}
  \qquad\mbox{where}\qquad
  \vr_{\{i\}}=\tr_{\{1,\ldots,n\}\setminus\{i\}}\vr.
\end{equation}
For the other projections there is no such explicit formula. 
Moreover, the family $\QQ_1$ is invariant under local 
filtering transformations of the form 
\be
\vr \mapsto 
\sigma
=
[F_1 \otimes F_2 \otimes ...\otimes F_n]
\vr
[F_1^\dagger \otimes F_2^\dagger \otimes ...\otimes F_n^\dagger]
\ee
where the $F_i$ are arbitrary matrices, since these transformations 
preserve the product structure \cite{verstraete}. This means that 
the quantity $D_1$ cannot increase under these transformations either 
\cite{GaGu11}. The exponential families $\QQ_k$ with $k\geq 2$
do not have this property.

\section{Symmetries}\label{sec:sym}
The computation of the quantities $D_k(\vr)$ and $C_k(\vr)$ is not 
straightforward. One may therefore ask, whether symmetries of 
the state $\vr$ can simplify the optimization procedure. For instance, if 
$\vr$ is invariant under the permutation of the first two 
particles, it seems natural that the state $\tau\in\QQ_k$ (and the 
corresponding $k$-party Hamiltonian) which minimize $D_k(\vr)$ 
share the same permutation symmetry. 
This symmetry assumption seems plausible, and the following Lemma 
presents a rigorous statement:
\\
{\bf Lemma 3.} Let $\vr$ be a quantum state which has a symmetry of
the form
\be
\vr = U \vr U^\dagger,
\ee
where $U$ is a unitary matrix that keeps the set of all $k$-particle
Hamiltonians invariant, i.e. it is a transformation so that the 
transformed operator $U H_k U^\dagger$ is again a $k$-particle 
Hamiltonian for any $k$-particle operator $H_k$. Then the state $\tilde{\vr}_k \in \QQ_k$  as well as the 
Hamiltonian $H_k$ minimizing the distance $D(\vr\Vert\tau)$ 
have the same symmetry as $\vr.$ This means that one can restrict 
the optimization to states and Hamiltonians which fulfil the condition
\be
\tau = U \tau U^\dagger \;\;\mbox{ and }\;\; H_k = U H_k U^\dagger,
\ee
respectively.

{\it Proof.} Consider a given $\vr$ with the symmetry and let $H \in \HH_k$ 
be the Hamiltonian of the state $\tau\in\QQ_k$ minimizing $D(\vr\Vert\tau).$
Here, $H$ denotes only the terms which are {\it not} proportional to the identity
in the Hamiltonian, i.e. one has $\tau = \exp(H)\exp[-\psi(H)],$ where $\exp[-\psi(H)]$
is the normalization [see Eq.~(\ref{eq:free})].
The idea is to consider $H'=(H+UHU^\dagger)/2$ and the corresponding 
$\tau'=\exp(H')$ and to prove that $D(\vr\Vert\tau')\leq 
D(\vr\Vert\tau).$ {From} the conditions on $U$ it then follows that $H' \in 
\HH_k$ and the uniqueness
of the information projection implies that $H=H'$ and therefore $H=UHU^\dagger.$

{From} Eq.~(\ref{ddarstellung1}) on sees
that $D(\vr\Vert\tau)$ and $D(\vr\Vert\tau')$ differ only in the contributions
from the normalizations $\psi(H)$ [or $\psi(H')$]. In fact, 
$D(\vr\Vert\tau')\leq D(\vr\Vert\tau)$
is equivalent to 
\be
\psi(H') = \ln\big\{\tr[\exp(H')]\big\} \leq \psi(H) = \ln\big\{\tr[\exp(H)]\big\}.
\ee
So it suffices to show $\tr[\exp(A+UAU^\dagger)] \leq \tr[\exp(2A)]$ for arbitrary
hermitean matrices $A.$ Applying the Golden-Thompson inequality 
$\tr[\exp(A+B)] \leq \tr[\exp(A)\exp(B)]$ (see page 261 in Ref.~\cite{bhatia}) 
we have
$\tr[\exp(A+UAU^\dagger)] \leq \tr[\exp(A) U \exp(A) U^\dagger]$
and it remains to show that
$\tr(X U X U^\dagger) \leq \tr(X^2)$
for $X=\exp(A).$ Taking the spectral decomposition $X=\sum_k \lambda_k \ketbra{\phi_k}$
this reads $\sum_{kl} C_{kl} \lambda_k \lambda_l \leq \sum_k\lambda_k^2,$
where $C_{kl}=|\bra{\phi_k} U \ket{\phi_l}|^2$ is a doubly stochastic matrix, 
that is, the row sums and column sums of $C$ equal one. Birkhoff's Theorem states that
any doubly stochastic matrix can always be written as a convex combination of permutation 
matrices, $C=\sum_k p_k \Pi_k$, where the $p_k$ form a probability distribution 
(see page 527 in Ref.~\cite{hornjohnson}). So it remains to show that $\sum_k \lambda_k \lambda_{\pi(k)} 
\leq \sum_k \lambda_k^2$ for an arbitrary permutation $\pi$, 
but this follows directly from the Cauchy Schwartz inequality.
\hfill $\square$

This Lemma can be used in various situations to simplify the calculation of the information 
projection:
\begin{itemize}

\item {\it Permutation symmetry:} If $\vr$ is invariant under permutation of the particles 
 $i$ and $j$, then this is a unitary symmetry as in Lemma 3, with the unitary flip operator
 $U = F_{ij}.$ This unitary operation also fulfils the other conditions of Lemma 3, and one can conclude
 that it suffices to optimize over Hamiltonians with the same permutation symmetry. Exploiting this symmetry can reduce the number of parameters in numerical algorithms significantly. 
 
 \item {\it Graph state symmetry:} In the framework of quantum information theory,
 so-called graph states and stabilizer operators have attracted significant 
 attention \cite{hein}. They are defined as follows: Consider an $n$-qubit system, 
 and $n$ observables $g_i$ which are tensor products of Pauli matrices  and which commute pairwise. For example, for three qubits one can take 
 $g_1 = \sigma_z \otimes \sigma_z \otimes \eins$,
 $g_2 = \eins \otimes \sigma_z \otimes \sigma_z$,
 and
 $g_3 = \sigma_x \otimes \sigma_x \otimes \sigma_x.$
 One can further consider all products of the $g_i.$ This is an the 
 Abelian group with $2^n$ elements (since $g_i^2=\eins)$ and this group
 is called the stabilizer. One can alternatively characterize the 
 stabilizer as all tensor products of Pauli matrices, which commute with
 all $g_i.$
 
 A graph state $\ket{G}$ is defined as an eigenstate of all $g_i$
 with eigenvalue $+1,$ that is $g_i\ket{G}= \ket{G}.$ For the 
 three-qubit example above, the graph state is given by the well-known
 GHZ state $\ket{GHZ}=(\ket{000}+\ket{111})/\sqrt{2}.$ Allowing also $\pm 1$
 eigenvalues one obtains a basis of $2^n$ graph states $\ket{G_k}.$
 
A quantum state diagonal in the basis of the $\ket{G_k}$ is a 
graph-diagonal state, and such states have been intensively studied 
\cite{duer,gjmw}. These states fulfil
 \be
 \vr_{\rm GD} = g_i (\vr_{\rm GD}) g_i^\dagger
 \ee
 and since the $g_i= g_i^\dagger$ are unitary, Lemma 3 can 
 be applied. One can directly see that the only possible 
 $k$-particle interaction terms in the Hamiltonian which 
 share the same symmetry are just all terms from the stabilizer 
 group which are of weight $k$ or less. For small $k$ these are 
 typically very few terms, which simplifies calculations significantly.
 Note that this structure was also observed for special graph-diagonal 
 states in Ref.~\cite{Zhou08}, but Lemma 3 shows that this holds for all mixed
 graph-diagonal states.
 
 \item {\it $U^{\otimes n}$-symmetry:} Another family of states 
 where Lemma 3 can be applied are the so-called $U^{\otimes n}$-invariant
 states. These states fulfil
 \be
 \vr =U^{\otimes n} \vr (U^\dagger)^{\otimes n}
 \ee
 for all possible unitary transformations $U$ on a single particle. For 
 two particles these states are the Werner states \cite{werner}, but 
 also for more particles detailed characterizations are known 
 \cite{eggeling, supersinglets}. Lemma 3 shows that when computing 
 $D_k(\vr)$ for these states, the optimal Hamiltonian has the same $
 U^{\otimes n}$-symmetry. This also implies that the optimal Hamiltonian
 has no single-particle terms, since the only single-qubit operator with
 this symmetry is the identity. It follows that for arbitrary
 $U^{\otimes n}$-invariant states the multiinformation is simply 
 given by
 \be
 D_1(\vr)= D_0(\vr)= n-S(\vr),
 \ee
 where $D_0(\vr)$ denotes the distance to the maximally mixed state.
 \end{itemize}

\section{Iterative computation of the quantum information projection}\label{sec:alg}
In this section we present an algorithm for computing the information 
projection. Other existing algorithms will be discussed in Section 5.
\label{sec:qcipi}

\subsection{Preliminary considerations}
In this section we will describe an efficient numerical algorithm 
with which to compute the projection $\tilde\varrho_k$ of a given 
$n$-particle quantum state $\varrho$ on an exponential family $\Q_k$. 
To this end one needs to construct $\tilde\varrho_k\in \Q_k$ such 
that $\tilde\varrho_k\in M_k(\varrho)$, i.e. such that $\tilde\varrho_k$ 
has the same $k$-particle reduced density matrices as $\varrho$. 
The algorithm we put forward is -- in spirit-- similar to the iterative 
projection algorithm for the classical case proposed in Refs.~\cite{CsSh04, cipi}.  
We start with the fully mixed state $\varrho'=\eins/\tr(\eins)$ as an 
approximation for $\tilde\varrho_k$, at each iteration step of the algorithm 
the current approximation of $\tilde\varrho_k$ is then improved such as to 
better match the reduced density matrix of $\vr$ defined by a particular subset 
of particles.  The algorithm proceeds by iteratively going through all
such subsets of at most $k$ particles repeatedly until convergence is reached.

Specifically, let $\vr$ be the state whose projection onto $\Q_k$ we want 
to calculate, and let $\tau=\e^H/\tr(\e^H)$ be the current approximation 
of $\tilde{\rho}_k$, where $H$ is a $k$-party Hamiltonian. The algorithm 
is initiated from $H=\eins$. 

For a fixed $\ell$-party observable $A$ with $\ell\le k$ the iteration step consists 
of adding a term $\varepsilon A$ to the Hamiltonian $H$, generating an updated 
approximation $\tau'$ of $\tilde\varrho$. The amplitude of the modification, 
$\varepsilon$, is chosen such that the expectation of $A$ under the density 
matrix $\tau'$ improves the match with the expectation obtained under $\varrho$. 
In detail one has the update
\begin{equation}
  \tau=\frac{\e^H}{\tr(\e^H)}
  \rightarrow
  \tau'(\varepsilon)=\frac{\e^{H+\varepsilon A}}{\tr(\e^{H+\varepsilon A})},
\end{equation}
  such that $\tr(A\tau')$ matches $\tr(A\varrho)$ as closely as possible. 
  In principle $\varepsilon$ can be obtained by solving $\tr(A\tau')=\tr(A\varrho)$ 
  directly. In practice this is hard to implement though, as $\tr(A\tau')$ depends 
  non-linearly on $\varepsilon$. We therefore resort to the following linear approximation
\begin{equation}
  \tr\left[A\tau'(\varepsilon)\right]=\tr(A\tau)
  +\varepsilon
  \left.\frac{\partial}{\partial\varepsilon}
  \tr\left[A\tau'(\varepsilon)\right]\right|_{\varepsilon=0}
  +{\cal O}(\varepsilon^2),
\end{equation}
where one has
\begin{equation}
 \frac{\partial}{\partial\varepsilon}\tr\left[A\tau'\right]
  =\tr\Bigl[A\frac{\partial_\varepsilon\e^{H+\varepsilon A}}
  {\tr(\e^{H+\varepsilon A})}\Bigr]
  -\tr\Bigl[A\frac{\e^{H+\varepsilon A}}{\tr(\e^{H+\varepsilon A})}\Bigr]
  \frac{\tr(\partial_\varepsilon\e^{H+\varepsilon A})}
  {\tr(\e^{H+\varepsilon A})}.\label{eq:alg1}
\end{equation}
We have here used the shorthand notation
$\partial_\varepsilon=\frac{\partial}{\partial\varepsilon}$. 
When evaluating the derivative of the matrix exponential 
in this expression one has to take into account that $H$ and $A$ 
generally do not commute. So it is convenient to use the 
identity \cite{Wilc67}
\begin{equation}
  \frac{\partial}{\partial t}\e^{M(t)}=\int_0^1~ ds~
  e^{sM(t)}\frac{\partial M(t)}{\partial t} e^{-sM(t)}e^{M(t)},
\end{equation}
valid for a general one-parameter family of matrices $M(t)$. 
Applying this identity to substitute for the derivatives 
in Eq. (\ref{eq:alg1}) and carrying out a modest amount of algebra one obtains
\be
\left.\frac{\partial}{\partial\varepsilon}
\tr\left[A\tau'(\varepsilon)\right]
\right|_{\varepsilon=0}
      =\frac{1}{\tr(\e^H)}\int_0^1 ds \tr\big[A\e^{sH}A\e^{-sH}\e^H\big]
    -\big[\tr(A\tau)\big]^2.
\ee
Next we apply a further approximation to the remaining integral, 
and replace it by the mean of the integrand evaluated at the upper 
and lower limits $s=0$ and $s=1$, respectively. This gives
\be
  \int_0^1ds\tr\bigl(A\e^{sH}A\e^{-sH}\e^H\bigr)
  \approx\frac{1}{2}\Bigl[\tr(A\e^HA)+\tr(A^2e^H)\Bigr]=\tr(A^2\e^H).
  \ee
This in turn leads to
\begin{equation}
  \tr[A\tau'(\varepsilon)]\approx\tr(A\tau)
  +\varepsilon\big\{\tr(A^2\tau)-[\tr(A\tau)]^2\big\}.
\end{equation}
Admittedly, this approximation can only be justified {\it a posteriori} 
by the performance of the algorithm. Setting $\tr[A\tau']=\tr[A\varrho]$ 
and using the approximation just obtained one finds the following 
solution for $\varepsilon$:
\begin{equation}
  \varepsilon\approx
  \frac{\tr(A\varrho)-\tr(A\tau)}{\tr(A^2\tau)-[\tr(A\tau)]^2}
  =\frac{\avg{A}_\varrho-\avg{A}_\tau}{\Delta^2_\tau(A)},
\end{equation}
where we have used the notation 
$\avg{A}_\varrho=\tr[A\varrho]$ (and 
analogously for $\avg{A}_\tau$), and where 
$\Delta^2_\tau(A)=\avg{A^2}_\tau-\avg{A}_\tau^2$ is the variance.

\subsection{Description of the algorithm}
In the full algorithm with which to compute the information projection 
of $\varrho$ onto the quantum exponential family $\Q_k$, one chooses 
an orthogonal
basis $V_k$ in the space of $k$-party observables (excluding the identity) and
updates the approximation $\tau$ for each $A\in V_k$ in turn. For an $n$-qubit
system one can
choose the Pauli operators 
\begin{equation}
  V_k=\bigl\{\tau_\alpha\bigm| 1\le W(\alpha)\le k\bigr\}.
\end{equation}

The complete algorithm is then as follows:\\

\noindent
\emph{Problem:} Given an $n$-qubit state $\varrho$, compute its information
  projection $\tilde{\varrho}_k$ onto the exponential family ${\Q}_k$.\\
  \noindent
  \emph{Algorithm:} 
  \begin{enumerate}
    \item[1.] Choose an orthonormal basis $V_k$ of the space of
      $k$-party observables, say these observables are $A_1, A_2,\dots, A_{M}$ (where $M$ will depend on $k$). For each element $A_i\in V_k$ compute the expectation value $\avg{A_i}_\varrho$.
    \item[2.] Initialize $\tau=\eins/2^n$ as the completely mixed state.
    \item[3.] \begin{enumerate} \item 
    	Start with $i=1$, and update $\tau$ according to 
      \begin{displaymath}
	\tau=\frac{\e^H}{\tr(\e^H)}
	\rightarrow
	\tau'=\frac{\e^{H+\varepsilon A_i}}{\tr(\e^{H+\varepsilon A_i})}
	\quad\mbox{where}\quad
	\varepsilon
	=\frac{\avg{A_i}_\varrho-\avg{A_i}_\tau}{\Delta^2_\tau(A_i)}.
      \end{displaymath}
    \item Increment $i$ to $i+1$ and repeat step 3(a). Once the coefficients 
    for all observable in $V_k$ have been updated (i.e for $i=1,\dots,M$) goto 4.
    \end{enumerate}
    \item [4.] If the maximum number of iterations has been reached or a 
    convergence criterion is met, terminate, otherwise goto 3.
  \end{enumerate}
When implementing the algorithm, it turns out to be useful to introduce an
additional parameter $\omega$ which controls the size of the steps in the space
of Hamiltonians,
\begin{equation}
  \tau=\frac{\e^H}{\tr(\e^H)}
  \rightarrow
  \tau'=\frac{\e^{H+\omega\varepsilon A}}
  {\tr(\e^{H+\omega\varepsilon A})}
\end{equation}
with $\varepsilon$ as above. Choosing values $\omega<1$ corresponds to what is known as a successive underrelaxation scheme \cite{numrec} and it can improve the convergence properties of the algorithm.

We would like to stress that we do not have a proof that the algorithm converges.
For the classical case, however, it has been shown that a similar algorithm converges 
\cite{CsSh04, cipi}. One could also think of improving our algorithm by using a better 
approximation to the integral. However, the numerical results shown below 
demonstrate that the algorithm as described here works remarkably well.

\subsection{Test of the algorithm}

In order to test the algorithm just described, we consider Dicke states of four and six qubits.
These states are given by
\begin{eqnarray}
\ket{D_{2}^4} &=& 
\frac{1}{\sqrt{6}} (\ket{0011}+\ket{0101}+\ket{0110}+ \ket{1001}+\ket{1010}+\ket{1100}),
\nonumber\\
\ket{D_{3}^6} &=& 
\frac{1}{\sqrt{20}}(\ket{000111}+\mbox{ permutations}),
\end{eqnarray}
that is, they are a balanced superposition of all terms in the standard 
basis with $n/2$ excitations. Dicke states have intensively been studied 
in entanglement theory and have been observed in several experiments \cite{tgreview}. 
As a one-parameter family of states, we consider Dicke states mixed with white 
noise, 
\begin{equation}
  \label{eq:whitenoisefam}
  \varrho^{(n)}(p)=p\frac{\eins}{2^n} +(1-p)\ketbra{D_{n/2}^n}.
\end{equation}
Before applying our algorithm, it is useful to discuss the symmetries of these 
states. First, the states $\varrho^{(n)}(p)$ are symmetric under the exchange 
of particles. Consequently, the information projection $\tilde\varrho_k$ shares
the same symmetry. There are, however, additional symmetries: If we consider 
the operators
\be
G^{(n)}_x=\sigma_x^{\otimes n} \mbox{ and } G^{(n)}_z=\sigma_z^{\otimes n},
\ee
then it directly follows that 
$G^{(n)}_\alpha [\varrho^{(n)}(p)] (G^{(n)}_\alpha)^\dagger = \varrho^{(n)}(p)$
for $\alpha = x,z$, which implies that the information projection $\tilde\varrho_k$ 
and the corresponding minimising Hamiltonians have this symmetry as well. Note that $\varrho^{(n)}(p)$ 
is also symmetric under the product $ G^{(n)}_y=G^{(n)}_x G^{(n)}_z$, but this is not 
an independent symmetry.

\begin{figure}
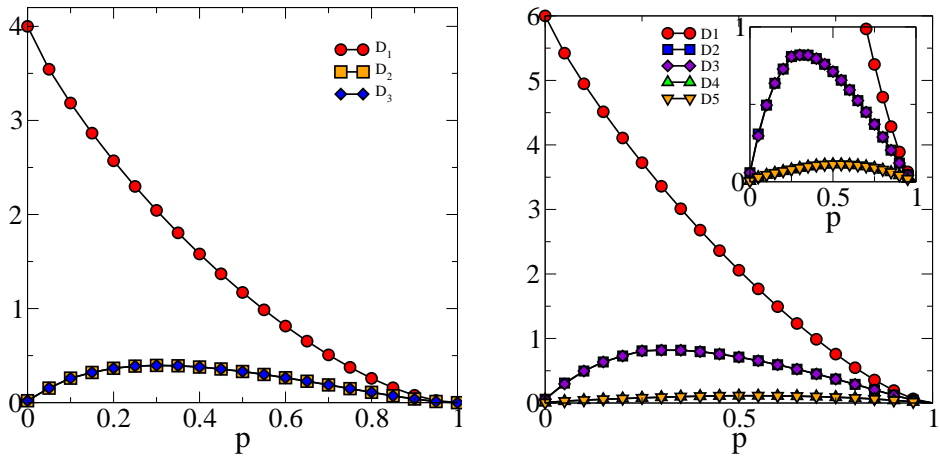

\vspace{0.7cm}
  \centering
  \includegraphics[height=6cm]{dicke4.eps}
  \hspace{0.5cm}
  \includegraphics[height=6cm]{dicke6.eps}
  \caption{\label{fig:dicke} Complexity measures for Dicke states
  mixed with white noise (left: the four-qubit case, right: the six-qubit case).
  See text for further details.
  }
\end{figure}

These symmetries reduce the number of parameters already significantly. For 
instance, there are no single-particle Hamiltonians, which are invariant under 
these symmetries, so we have $D_1(\vr)= D_0(\vr)= n-S(\vr)$, as for the 
$U^{\otimes n}$-invariant states discussed above. Similarly many possible interaction terms can be discarded from the outset for higher-order interactions.  Applying our algorithm generates the data shown in Fig.~\ref{fig:dicke}. We have here chosen an under-relaxation parameter of $\omega=0.5$ for $\varrho^{(4)}(p)$ [$\omega=0.1$ for $\varrho^{(6)}(p)$], although $\omega=1$ yields similar results when convergence is reached. We typically run the algorithmic scheme for up to $100$ iterations [$500$ iterations for  $\varrho^{(6)}(p)$], or until a convergence threshold is met, and we report the minimum distance reached over this number of iterations. It must be stressed though that our numerical results are estimates of the respective distances, we cannot exclude that the precise quantitative results have a remaining dependence on parameters of the algorithm, such as the relaxation parameter $\omega$, the maximum number of iteration steps, or the precise convergence criterion.

For the four-qubit case, one finds that the measures $D_2$ and $D_3$ coincide.
This can be explained as follows: Let us consider the information projection 
$\tilde \vr_2=\exp(H_2)$ and the corresponding two-particle Hamiltonian 
$H_2 \in \HH_2.$ As already mentioned above, $H_2$ does not contain any 
single-particle term. For the two-particle terms there are also not many 
possibilities, in fact, the only possible terms are 
$h_\alpha^{1,2} = \sigma_\alpha \otimes \sigma_\alpha \otimes \eins \otimes 
\eins$ for $\alpha=x,y,z$, and permutations thereof. Using the power series of 
the  exponential function, one finds that $\tilde \vr_2=\exp(H_2)$ has no 
three-body correlations in its Bloch representation, that is, 
$\tr[(\sigma_i \otimes \sigma_j \otimes \sigma_k \otimes \eins) \tilde \vr_2]=0$ 
for any choice of 
$i,j,k \in \{x,y,z\}.$\footnote{The detailed proof is the following: 
$H_2$ and $\tilde \vr_2$ obey the symmetry defined by the $G^{(n)}_\alpha$,
which implies already that most of the three-body correlations in the Bloch 
representation vanish. The only terms which are not forced to be zero are 
expectation values of $K^{(3)}= \sigma_x \otimes \sigma_y \otimes \sigma_z \otimes \eins$
or permutations thereof. However, if $\exp(H_2)$ is written as a power series,
the term $K^{(3)}$ does never occur as a product of the two-qubit terms $h_\alpha^{i,j}.$
The reason is that if the product of the single-qubit observables in $K^{(3)}$ is taken, 
the result is $(\sigma_x)\cdot(\sigma_y)\cdot(\sigma_z)\cdot(\eins) = i \eins,$ 
which is a non-hermitian operator. If an arbitrary product of the $h_\alpha^{i,j}$
is considered, and then the product of the single-qubit observables is taken, the result 
is an hermitean operator, since any Pauli matrix occurs an even number of times in the 
total product.}
In other words, one can state that the maximizer of the entropy in the linear 
family $\MM_2$ has no three-body correlations in its Bloch representation.

On the other hand, let us consider the three-body reduced state of 
$\varrho^{(4)}(p).$ This state is given by 
\be
\vr_{123}=p \frac{\eins}{2^3} + \frac{1-p}{6}(\eins-\ketbra{000}-\ketbra{111})
\ee
and one can directly check that this state also has no three-body correlations
in its Bloch representation. But this means that the maximizer of the the entropy 
in the linear family $\MM_2$ has the same reduced three-particle density matrices
as $\varrho^{(4)}(p)$, so it is also an element of the smaller linear family 
$\MM_3.$ From this $D_2[\varrho^{(4)}(p)]=D_3[\varrho^{(4)}(p)]$ follows.

For the six-qubit case, we find that $D_2[\varrho^{(6)}(p)] = D_3[\varrho^{(6)}(p)]$
and $D_4[\varrho^{(6)}(p)] = D_5[\varrho^{(6)}(p)]$ and this can be understood
in the same way. In order show that $D_2=D_3$ one first finds that the reduced
three-particle state of $\varrho^{(6)}(p)$ does not contain any three body 
correlations, then the argument is the same as for the four-qubit Dicke state. 
For $D_4=D_5$ one has to consider the reduced five-particle density matrices.
Due to the symmetry, the only relevant correlators are 
$K^{(5)}_1=\sigma_x \otimes \sigma_z \otimes \sigma_y \otimes \sigma_y 
\otimes \sigma_y\otimes \eins$,
$K^{(5)}_2=\sigma_x \otimes \sigma_z \otimes \sigma_z \otimes \sigma_z 
\otimes \sigma_y\otimes \eins$
and
$K^{(5)}_3=\sigma_x \otimes \sigma_x \otimes \sigma_x \otimes \sigma_z 
\otimes \sigma_y\otimes \eins.$ Their mean values vanish in the 
state $\varrho^{(6)}(p),$ and then the proof can proceed as before.

Besides these examples, we have tested our algorithm for a variety of four- 
and five-qubit states, and previous results \cite{Zhou08, Zhou09b} we easily
reproduced. This shows that the algorithm presented above is a useful tool
for computing the complexity measure for states up to six qubits.

\section{Other algorithms}
In this section, we will first describe another algorithm, 
which has been proposed to compute the complexity measure 
\cite{Zhou09b}. Then, we will discuss how other methods known
in numerical optimization can be used for this problem.

\subsection{The algorithm of Ref.~\cite{Zhou09b}}
In Ref.~\cite{Zhou09b} D.L. Zhou has proposed an algorithm
for computing the information projection and has presented 
examples up to five qubits. The idea of the algorithm is 
as follows.

First, one considers the information projection $\tilde\vr_k$
onto the exponential family $\QQ_k$ and its logarithm 
$\log(\tilde\vr_k)$, which is effectively the generating 
$k$-particle Hamiltonian. Then, according to Lemma 2, 
$\tilde\vr_k$ obeys the following conditions: (i) First, 
the mean values of Pauli matrices $\sigma_\alpha$ in the state 
$\tilde\vr_k$ equal the mean values in the state $\vr$, 
if the weight $W(\alpha)$ of $\sigma_\alpha$ is smaller or equal
$k.$ This is nothing but the condition that the reduced $k$-particle
states of $\tilde\vr_k$ and $\vr$ are the same. (ii) Second, the mean
values of Pauli matrices $\sigma_\alpha$ in the Hamiltonian 
$\log(\tilde\vr_k)$ vanish, if the weight $W(\alpha)$ of 
$\sigma_\alpha$ is larger than $k.$ This is the condition that 
the generating Hamiltonian contains no higher-order interactions than
$k$-particle interactions.

These conditions lead to $4^n$ equalities for $\tilde\vr_k$, and 
from Lemma 2 it follows that there is a unique solution to all these 
equalities, which is the desired $\tilde\vr_k$. Due to the occurrence 
of the logarithm, however, the equalities are highly nonlinear, and a 
direct numerical solution is not straightforward. Therefore, as 
explicitly stated in Ref.~\cite{Zhou09b}, one needs an initial 
guess of an initial value of $\tilde\vr_k$ for solving them. For 
example, in Ref.~\cite{Zhou09b} curves like the ones in Fig.~2 
have been computed iteratively: Initially, one solves the problem 
if the state is completely noisy ($p=1$), then one uses the solution
as an initial value for solving the nonlinear equations for decreasing
$p \mapsto p-\varepsilon$ and so on, until $p=0$ is reached. We stress that no such procedure is needed for our algorithm, then data points shown in Fig. \ref{fig:dicke} are obtained independently for the different values of $p$.

\subsection{Convex optimization approaches}

The algorithm described in Section~4 searches for an
approximation of the information projection by an iteration
within the exponential family $\mathcal Q_k$, which is a highly 
nonlinear manifold. In Lemma~2 it was established that the 
information projection $\tilde \varrho_k$ can also be characterised 
by a maximization of the von Neumann entropy $S(\varrho')$ over the 
linear family $\mathcal M_k(\varrho)$, given by the density matrices 
with the same $k$-particle reduces density matrices. The advantage of 
this formulation is that the problem becomes an instance of convex 
optimization, namely the minimization of the convex function 
$-S(\varrho')$ over the convex set $\mathcal M_k(\varrho)$.
Note that convex optimization problems are well-studied, for 
an overview see Ref.~\cite{boyd}.

{From} this structure it is clear that no local minima exist, 
i.e., if $-S(\varrho')\le -S(\tau)$ for all $\tau\in \mathcal M_k(\varrho)$ 
with $\|\varrho'-\tau \| < \varepsilon$ where $\varepsilon>0$, 
then $\varrho'$ attains the global minimum. This makes a numerical 
solution particularly tractable.  In the language of 
convex optimization, the problem reads
\begin{eqnarray}
\mbox{maximize: } && t 
\nonumber
\\
\mbox{subject to: } && S(\varrho')\ge t,
\nonumber\\
                  && \tr[(\varrho'-\varrho) \sigma_\alpha]= 0
\mbox{ for all } \alpha\mbox{ with } W(\alpha)\le k \mbox{, and}
\nonumber\\
 && \varrho'\ge 0.
\end{eqnarray}
For such problems one can construct algorithms where the optimality 
of the solution is guaranteed.

A problem which is related to the one discussed here was studied 
by Teo and coworkers in the context of quantum state 
estimation theory \cite{TZER11}: They consider the situation 
where in a quantum experiment the observed frequencies $\mathbf f$ 
of measurement outcomes (described by positive operators $E_i$)
are sampled from a probability distribution $P[\varrho']$ from an 
unknown quantum state $\varrho'$ (that is, the probabilities are 
computed according to $P_i = \tr(E_i \vr')$). 
Among the solutions which maximize the log-likelihood 
$\log(L(\varrho'|\mathbf f))$ [which is proportional to $-D(\mathbf f\|P[\varrho'])$], 
they propose to use the state with maximum entropy. For the solution of 
this problem they introduce an iterative algorithm that eventually 
approaches
\begin{equation}\label{eq:tao}
 \varrho^* = \lim_{\lambda\searrow 0} \argmin_{\varrho'} [-\log(
 L(\varrho'|\mathbf f))-\lambda S(\varrho')],
\end{equation}
 where the minimization is performed over all $\varrho'\ge 0$.

 In order to see the relation to our problem, consider a probability
distribution $P[\tau]_i = \tr(E_i \tau)$, where $E_i$ are positive
semidefinite operators with $\sum_i E_i=\eins$ and $\mbox{span} \{E_i \} =
\mbox{span}\, \mathcal M_k(\varrho)$. This means that knowledge of
the probabilities $P[\tau]$ is equivalent to knowledge of the 
reduced $k$-particle density matrices. If we now replace $\mathbf f$ 
by $P[\varrho]$, then the optimization in Eq.~(\ref{eq:tao}) is indeed 
closely related to our optimization problem: One can view Eq.~(\ref{eq:tao})
as a maximization of the entropy $S(\varrho')$ where the log-likelihood 
term [being proportional to $D(P[\varrho]\|P[\varrho'])$], 
serves as a barrier term, forcing $\varrho$ and $\varrho'$ to have
the same reduced $k$-particle states. Such constructions are known from 
interior point methods for convex optimization \cite{boyd}.

While the methods presented in this section rely on established 
numerical algorithms, we observed that computing the information 
projection by solving a convex optimisation problem actually requires more 
resources than the algorithm we propose in Sec.~4. This is probably
due to the fact, that the algorithm of Sec.~4 exploits the structure 
of the problem in a better way.

\section{Conclusions}
In summary we have used concepts from information geometry 
to characterise the complexity of multiparticle quantum states. 
Specifically, we considered the distance of a given $n$-particle 
density matrix from  the space of all thermal quantum states 
generated by Hamiltonians with $k$-particle interactions. 
We have shown how symmetries can be used to simplify the 
calculation of the resulting complexity measure. Furthermore, 
we have proposed a new algorithm to compute this measure. This algorithm, 
we think, is computationally more efficient than existing approaches, 
and in particular we are able to compute the above complexity measures 
for selected six-particle states.

There are several follow-on problems requiring further attention. First, 
the complexity measure is not yet fully understood, and several interesting
open questions remain, for example, which are the states with a maximal distance $D_k$ from a given exponential family? How is the 
complexity measure related to known entanglement measures or correlation 
measures? Second, it would be interesting to study this measure in 
specific situations. For instance, for a given $n$-particle spin model with two-particle 
interactions only one may consider the reduced states of some of the particles and 
ask, whether they still can effectively be described as thermal states
of a two-body Hamiltonian, or whether higher-order correlations are present. We expect that this may for example be of interest in models of quantum spin chains at or near quantum phase transitions.  It is known that entanglement measures can be used to
study such critical phenomena, and so a systematic exploration of the complexity measures we have proposed here cannot only help to understand quantum phase transitions better, but also to relate different measures of complexity and correlation.


\ack{We thank Tobias Moroder for helpful discussions. This work is 
supported by the Royal Society (reference JP090467), the  EU (Marie Curie 
CIG 293993/ENFOQI) and the BMBF (Chist-Era Project QUASAR).}

\section*{References}

\begin{thebibliography}{99}

\bibitem{plenio07}
M. B. Plenio and S. Virmani,
Quant. Inf. Comp. {\bf 7}, 1 (2007).

\bibitem{hororeview}
R. Horodecki, P. Horodecki, M. Horodecki, and K. Horodecki,
{Rev. Mod. Phys.} {\bf 81}, 865 (2009).

\bibitem{tgreview}
O. G\"uhne and G. T\'oth,
{Phys. Rep.} {\bf 474}, 1 (2009).

\bibitem{cumulant}
D. L. Zhou, B. Zeng, Z. Xu, and L. You,
Phys. Rev. A {\bf 74}, 052110 (2006).

\bibitem{discordreview}
K. Modi, A. Brodutch, H. Cable, T. Paterek, and V. Vedral,
Rev. Mod. Phys. {\bf 84}, 1655 (2012).

\bibitem{complexity1}
R. Badii and A. Politi,
{\it Complexity: Hierarchical Structures and Scaling in Physics},
Cambridge University Press (1999).

\bibitem{complexity2}
K. Lindgren and M. G. Nordahl, 
Complex Systems {\bf 2}, 409 (1988).

\bibitem{complexity3}
S. Lloyd, Control Systems, IEEE {\bf 21} 7 (2001)

\bibitem{complexity4}
D. P. Feldman and J. P. Crutchfield, 
{\em A survey of complexity measures}, 
Santa Fe Summer School 1998

\bibitem{complexity5} 
D. Feldman and J. P. Crutchfield, 
Phys. Lett. A {\bf 238}, 244 (1998).

\bibitem{complexity6} 
B. Edmonds, 
{\em Bibliography of Complexity Measures}, 
available at {\tt bruce.edmonds.name/combib/}.

\bibitem{KOJA09}
T.~Kahle, E.~Olbrich, J.~Jost and N.~Ay, 
Phys. Rev. E \textbf{79}, 026201 (2009).

\bibitem{Zhou08}
D.~L. Zhou, 
Phys. Rev. Lett. \textbf{101}, 180505 (2008).

\bibitem{Zhou09}
D.~L. Zhou, 
Phys. Rev. A \textbf{80}, 022113 (2009).

\bibitem{GaGu11}
T.~Galla and O.~G\"uhne, 
Phys. Rev. E {\bf 85}, 046209 (2012). 

\bibitem{amari01}
S. Amari, 
IEEE Trans. Inf. Theory {\bf 47}, 1701 (2001).

\bibitem{CsSh04}
I.~Csisz\'ar and P.~C. Shields, 
{Found. and Trends in
  Communications and Inf. Theory \textbf{1}, 417} (2004).

\bibitem{cipi}
L. Steiner and T. Kahle
{\it Computing Information Projections Iteratively with {CIPI}},
available at {\tt github.com/tom111/cipi}, (accessed August 2010).

\bibitem{Zhou09b}
D.~L. Zhou, 
arXiv:0909.3700.

\bibitem{AyKnauf07}
N. Ay and A. Knauf, 
Kybernetika {\bf 42}, 517 (2007).

\bibitem{stingl} 
M. Stingl, {\it Statistische Physik}, lecture notes at the Universit\"at M\"unster, 
available at
{\tt pauli.uni-muenster.de/tp/fileadmin/lehre/skripte/stingl/spskript.ps.zip}. 

\bibitem{zia}
R. K. P. Zia, E. F. Redish and S. R. McKay, 
American Journal of Physics {\bf 77}, 614 (2009).
 
 \bibitem{balian} 
 R. Balian, 
 {\em Du microscopique au macroscopique: cours de physique statistique de l'Ecole polytechnique}, 
 Ellipses-Edition Marketing, Paris 1982
 
\bibitem{kimura}
G. Kimura,
Phys. Lett. A {\bf 314}, 339 (2003).

\bibitem{byrd}
M.~S. Byrd and N. Khaneja,
Phys. Rev. A {\bf 68}, 062322 (2003).


\bibitem{verstraete}
F. Verstraete, J. Dehaene, and B. De Moor,
Phys. Rev. A {\bf 68}, 012103 (2003).


\bibitem{bhatia}
R. Bhatia, 
{\it Matrix Analysis}, 
Springer-Verlag (Heidelberg) 1997.

\bibitem{hornjohnson}
R. Horn and C. Johnson, 
{\it Matrix Analysis},
Cambridge University Press (Cambrigde) 1999.

\bibitem{hein}
M. Hein, W. D\"ur, J. Eisert, R. Raussendorf, M. Van den Nest,  and 
H.-J. Briegel, {\it Entanglement in Graph States and its Applications},
Proceedings of the International School of Physics "Enrico Fermi" on 
"Quantum Computers, Algorithms and Chaos", Varenna, Italy (2005);  
quant-ph/0602096.

\bibitem{duer}
M. Hein, W. D\"ur, and H.-J. Briegel,
Phys. Rev. A {\bf 71}, 032350 (2005).

\bibitem{gjmw}
O. G\"uhne, B. Jungnitsch, T. Moroder, and Y.S. Weinstein,
Phys. Rev. A {\bf 84}, 052319 (2011). 

\bibitem{werner}
R.~F. Werner, 
Phys. Rev. A {\bf 40}, 4277 (1989). 

\bibitem{eggeling}
T. Eggeling and R.~F. Werner, 
Phys. Rev. A {\bf 63}, 042111 (2001). 

\bibitem{supersinglets}
A. Cabello, 
J. Mod. Opt. {\bf 50}, 10049 (2003).

\bibitem{Wilc67}
R.~M. Wilcox, 
{J. Math. Phys. \textbf{8}, 962} (1967).

\bibitem{numrec} 
W. H. Press, S. A. Teukolsky, W. T. Vetterling, B. P. Flannery, 
{\em Numerical Recipes 3rd Edition: The Art of Scientific 
Computing}, Cambridge University Press, New York (2007)

\bibitem{boyd}
S. Boyd and L. Vandenberghe, 
{\it Convex Optimization},
Cambridge University Press 2004.

\bibitem{TZER11}
Y.~S. Teo, H.~Zhu, B.-G. Englert, J.~\v{R}eh\'a\v{c}ek and Z.~Hradil,
{Phys. Rev. Lett. \textbf{107}, 020404} (2011).




\end{thebibliography}

\end{document}